\documentclass{optica-article}

\journal{opticajournal} 

\articletype{Research Article}
\usepackage[title]{appendix}%
\usepackage{hyperref}

\begin{document}

\title{Scattering Correction through Fourier-Domain Intensity Coupling in Two-Photon Microscopy (2P-FOCUS)}

\author{Daniel Zepeda,\authormark{1} Yucheng Li,\authormark{1} and Yi Xue\authormark{1,*}}

\address{\authormark{1}Department of Biomedical Engineering, University of California, Davis, 451 Health Sciences Dr., Davis, CA, 95616, USA}

\email{\authormark{*}yxxue@ucdavis.edu} 


\begin{abstract*} 
Light penetration depth in biological tissue is limited by tissue scattering. Correcting scattering becomes particularly challenging in scenarios with limited photon availability and when access to the transmission side of the scattering tissue is not possible. Here, we introduce a new two-photon microscopy system with Fourier-domain intensity coupling for scattering correction (2P-FOCUS). 2P-FOCUS corrects scattering by intensity modulation in the Fourier domain, leveraging the nonlinearity of multiple-beam interference and two-photon excitation, eliminating the need for a guide star, iterative optimization, or measuring transmission or reflection matrices. 2P-FOCUS uses random patterns to probe scattering properties, combined with a single-shot algorithm to rapidly generate the correction mask. 2P-FOCUS can also correct scattering beyond the limitation of the memory effect by automatically customizing correction masks for each subregion in a large field-of-view. We provide several proof-of-principle demonstrations here, including focusing and imaging through a bone sample, and imaging neurons and cerebral blood vessels in the mouse brain ex vivo. 2P-FOCUS significantly enhances two-photon fluorescence signals by several tens of folds compared to cases without scattering correction at the same excitation power. 2P-FOCUS can also correct tissue scattering over a 230$\times$230$\times$510 $\mu$m\textsuperscript{3} volume, which is beyond the memory effect range. 2P-FOCUS is able to measure, calculate, and correct scattering within a few seconds, effectively delivering more light deep into the scattering tissue. 2P-FOCUS could be broadly adopted for deep tissue imaging owing to its powerful combination of effectiveness, speed, and cost.

\end{abstract*}

\section{Introduction}\label{sec1}
Noninvasive focusing of light and imaging of objects embedded in scattering tissue are crucial for both basic biological research and clinical applications. Tissue scattering is the primary limitation for deep tissue focusing and imaging. Advanced microscopy techniques, such as confocal microscopy and multiphoton microscopy, enable noninvasive deep tissue imaging. However, as imaging depth increases, the number of ballistic photons available for excitation at a target location decreases exponentially. Eventually, the background intensity becomes comparable to the signal in non-sparsely labeled samples \cite{Wang2020-zb}.

To increase fluorescence signal intensity in two-photon microscopy, adaptive optics has been employed to correct low-order aberrations \cite{Ji2017-vf, Booth2007-vo}. Recent cutting-edge research \cite{Streich2021-wf, Rodriguez2021-sy} demonstrates that combining adaptive optics with multiphoton microscopy can enhance the fluorescence signal by several folds. A spatial light modulator (SLM) or a deformable mirror, positioned at the conjugate plane of the objective lens's back aperture, is used to modulate the phase of light in the Fourier domain. However, the patterning speed of SLM is relatively slow ($<$1 kHz), limiting its application to correct dynamic scattering \cite{Qureshi2017-on}.

Numerous techniques have been developed to correct highly spatially varying scattering \cite{Gigan2022-bc, Yoon2020-kg, Horstmeyer2015-yw, Mosk2012-lc, Vellekoop2015-wz}. However, most of these techniques are designed for transmission mode, which limits their applications for imaging objects embedded within the scattering tissues of living animals. Among the reflection mode techniques, many are post-processing methods \cite{Kang2017-hn, Yoon2020-sm, Boniface2020-yf, Wijethilake2023-ym, Escobet-Montalban2018-zc}, including our previous work \cite{Xue2018-ay, Xue2022-rf, Xue2024-aa, Zheng2021-pm, Wei2023-km}, which require significant time for image reconstruction ranging from several minutes to over an hour. Unlike post-processing techniques that correct scattering in images offline, several techniques achieve active correction of scattering by wavefront shaping using an SLM in reflection-mode multiphoton microscopy systems \cite{Tang2012-bl, Papadopoulos2016-sy, May2021-er, Qin2022-ns}. The wavefront shaping is based on interferometry measurements. Excitation light is split into two beams, either in a common path \cite{Tang2012-bl, May2021-er} or in a divergent path \cite{Papadopoulos2016-sy, Qin2022-ns}. One beam is steady, while the other beam is phase-shifted using either an SLM or a galvo-scanner. The interference of these two beams changes with the phase shifts, resulting in changes in fluorescence intensity, which is used as the feedback signal. The phase mask can then be calculated and optimized after a few iterations. Even though these techniques achieve deep tissue imaging, they still face some limitations. First, the maximum improvement in fluorescence intensity for brain imaging is usually less than tenfold \cite{Papadopoulos2016-sy, May2021-er, Qin2022-ns}. Second, the field-of-view after scattering correction is usually limited to a few tens of microns in brain imaging due to the limited range of the memory effect (i.e., isoplanatic patch, mean free path). Third, the total correction speed is relatively slow, including not only the time for taking measurements but also the time for computing the mask, projecting the mask, and reloading updated phase patterns to the SLM for iterative optimization. A detailed comparison between these methods and our technique is summarized in Appendix Table \hyperlink{appendixtable}{\ref*{table1}}.

Active correction of scattering via intensity modulation using a digital micromirror device (DMD) enables high-speed scattering correction. ``Intensity modulation” here refers to techniques that directly modulate the intensity of light at the Fourier plane as opposed to ``phase modulation” discussed above which directly modulates the phase of light at the Fourier plane. Most commercially available DMDs have patterning speeds ranging from tens to over 30 kHz, which is significantly faster than the state-of-the-art SLMs. DMDs are also generally more cost-effective than SLMs. Intensity modulation for scattering correction has been demonstrated by focusing a laser beam through a scattering medium and detecting the transmitted laser light with a detector \cite{Akbulut2011-zt, Conkey2012-st, Zhao2021-wg, Dremeau2015-fy, Zhang2014-ah, Nam2020-nv, Yang2021-ug}. The correction mask can be probed by turning on segments on the intensity modulator one-by-one \cite{Akbulut2011-zt}, by projecting Hadamard patterns to calculate transmission matrix \cite{Conkey2012-st, Zhao2021-wg}, or by projecting random patterns combined with transmission matrix measurement \cite{Dremeau2015-fy} or iterative optmization algorithms \cite{Zhang2014-ah, Nam2020-nv, Yang2021-ug}. Compared to phase modulation, traditional intensity modulation faces a challenge: for a given incident laser power on the modulator, scattering correction increases the signal, but intensity modulation reduces the signal because the total intensity on the detector decreases with the turned-off pixels. This is not an issue when focusing a laser beam through scattering media and detect the transmitted laser light \cite{Akbulut2011-zt, Conkey2012-st, Zhao2021-wg, Dremeau2015-fy, Zhang2014-ah, Nam2020-nv, Yang2021-ug}, as the photon flux from a laser is high. 

However, in fluorescence imaging, especially multiphoton imaging of biological samples, scattering correction by intensity modulation is very challenging because: (1) The number of fluorescent photons under two-photon excitation is much lower than that of a laser beam, and the maximum excitation intensity is constrained by the thermal damage threshold of biological samples; (2) Fluorescent photons are collected in reflection mode rather than in transmission mode, while most light is forward-scattered by biological tissues, further reducing the number of photons after scattering; (3) Two-photon microscopy images embedded objects within scattering tissue, without any free space between the phantom and the focus, which limits the degrees of freedom for correcting scattering \cite{Katz2012-oy, Jeong2018-lb}; (4) Imaging a large field of view requires scattering correction beyond the memory effect range, which is more difficult than correcting scattering at a single spot. 

Recently, high-power femtosecond lasers (ranging from a few watts to over 100 W) for multiphoton microscopy have become commercially available, making two-photon microscopy with intensity modulation feasible \cite{Ren2020-xy, Chen2020-qv}. It has been used for aberration correction \cite{Ren2020-xy, Chen2020-qv}, but not yet for scattering correction because the challenges mentioned above remain unsolved. In these work \cite{Ren2020-xy, Chen2020-qv} for aberration correction, the DMD is used as a wavefront modulator by calculating binary holographic masks using Lee holography \cite{Lee1979-rd}, with Zernike modes employed as the orthogonal basis to probe aberrations. The correction mask is generated using a modified hill-climbing algorithm similar to the classic algorithm in adaptive optics \cite{Booth2007-op}, where the final correction mask is a linear combination of weighted Zernike modes, and the weight coefficient of each Zernike mode is iteratively measured. However, binary intensity modulation generated by Lee holography is less efficient and effective than phase modulation, which has been demonstrated for focusing through scattering media \cite{Kim2014-zq,Dremeau2015-fy}. Consequently, while this technique corrects aberrations in two-photon microscopy, fluorescence intensity is only improved by 2.5 to 4 times when imaging ex vivo Drosophila brains at depths of 100 - 200 $\mu$m. Despite the DMD's fast operational speed of 22.7 kHz, the iterative hill-climbing algorithm is slow, leading to a total processing time of approximately 6 minutes. Therefore, instead of adapting orthogonal bases (i.e., Zernike modes) and algorithms (i.e., iterative hill-climbing algorithm) designed for phase modulation, innovative approaches \textit{tailored specifically for two-photon microscopy with binary intensity modulation} are needed, including new orthogonal bases and algorithms. While the key advantage of DMD-based intensity modulation is its patterning speed, no two-photon microscopy system with binary intensity modulation has yet been able to rapidly and effectively correct scattering in deep tissue. 

To overcome these challenges, we have developed a new two-photon microscopy system named 2P-FOCUS (\underline{FO}urier-domain intensity \underline{C}o\underline{u}pling for \underline{S}cattering correction) to provide effective and fast scattering correction over a large field of view. The power loss from intensity modulation is compensated by increasing the laser power input to the intensity modulator (i.e., a DMD) to maintain the same laser power on the sample before and after modulation. Unlike phase modulation, intensity modulation segments the incident collimated beam into multiple narrow beams, generating multiple-beam interference within the scattering tissue. To select the beams that remain in phase after scattering while blocking those that are out of phase, we use random patterns to probe whether the randomly selected beams are in phase or not after scattering. In-phase beams generate a bright main lobe after interference, resulting in brighter fluorescence compared to non-constructive interference. The final correction mask is generated by summing those random patterns followed by binarization. Random patterns are used instead of patterns at specific frequencies (e.g., Hadamard \cite{Conkey2012-st, Zhao2021-wg} or Zernike patterns \cite{Ren2020-xy, Chen2020-qv}) because: (1) Each random pattern consists of a broad range of spatial frequencies, allowing us to multiplex measurements across different frequencies; (2) random patterns can flexibly control the ratio of turned-on pixels to the total number of pixels (``sparsity") to improve the correction but Hadamard or Zernike patterns cannot; and (3) random patterns generate a central focus with stable intensity and location, whereas Hadamard patterns produce multiple foci that vary in intensity and position with each pattern. This single-shot algorithm ensures constructive interference without the need for time-consuming iterative optimization. To correct scattering beyond the memory effect range, we synchronize the DMD with a galvo-scanner on its conjugate plane to project different correction masks onto corresponding regions during scanning. This strategy significantly increases the system's degrees of freedom and effectively controlling a 4D light field \cite{Xue2022-yr}. 

We demonstrate the impressive capabilities of 2P-FOCUS through a series of proof-of-principle experiments. These include focusing and imaging through approximately 200 $\mu$m-thick bone and capturing fluorescence-labeled neurons and blood vessels at depths of up to 510 $\mu$m deep in ex vivo mouse brain tissue. Remarkably, 2P-FOCUS achieves over a 36-fold increase in fluorescence intensity compared to standard two-photon microscopy when imaging in the brain tissue. 2P-FOCUS also achieves a 230 $\times$ 230 $\mu m^2$ field-of-view in all experiments---beyond the memory effect range and significantly larger than that of current methods. Furthermore, 2P-FOCUS completes the \textit{entire} correction process in just \textit{a few seconds}, including the time for measurements, calculation, and projection of correction masks, far surpassing the speed of similar techniques \cite{Ren2020-xy, Chen2020-qv}. To our knowledge, 2P-FOCUS is the first two-photon microscopy system to use intensity modulation for rapid and effective scattering correction. 

\section{Methods}\label{sec2}
\subsection{Principle}\label{subsec21}
\begin{figure}[ht]%
\centering
\includegraphics[width=0.82\textwidth]{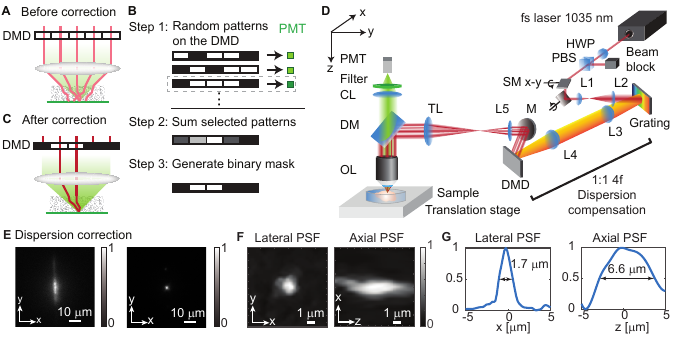}
\caption{Principle and schematic diagram of 2P-FOCUS. (A) Without correction, all pixels of the digital micromirror device (DMD) are turned on, the incident light is scattered and cannot form a tight focus. (B) The process of generating a correction mask involves three steps. (C) With correction, the binary correction mask is projected on the DMD, allowing only the beams that interfere constructively pass through. The illumination power on the sample is maintained the same before and after correction by increasing the input power to the DMD to compensate for the power loss due to turning off some pixels. The correction results in a brighter focus compared to the case before correction. (D) Optical schematic diagram of 2P-FOCUS. Details are in Appendix~\ref{setup}. (E) The lateral point-spread-function (PSF) before (left) and after (right) dispersion compensation, measured by a camera in reflection mode. (F) The lateral (left) and axial (right) PSF without scattering media, measured by the PMT in (D). (G) Intensity profile of the PSFs in (F). The lateral resolution is 1.7 $\mu$m and the axial resolution is 6.6 $\mu$m without scattering media.}\label{fig1}
\end{figure}

2P-FOCUS corrects scattering in three steps: taking measurements, calculating the correction mask, and projecting the correction mask (Fig.~\ref{fig1}A-C). First, 2P-FOCUS projects random intensity patterns in the Fourier domain while monitoring the fluorescence intensity excited under this modulation (Fig.~\ref{fig1}B, step 1). In the following experiments, random patterns are generated from a uniform discrete distribution on [0, 1] using MATLAB, with the random seed shuffled each time a new set of random patterns is created. Different sets of random patterns, consisting of hundreds to thousands of patterns, are generated for varying sparsity levels and super-pixel sizes. Random intensity patterns has been used as an orthogonal basis in computational optics \cite{Baraniuk2007-wm}. Random patterns are orthogonal to each other because each of them consists of a large number of segments (hundreds to thousands), thus satisfying the mathematical criteria of an orthogonal matrix, as $P^TP=I$, where $P$ is the reshaped and normalized random pattern matrix and $I$ is the identity matrix. In other words, the dot product of any two distinct random patterns is approximately zero. Each segment on the DMD consists of $n\times n$ pixels, termed ``super-pixels”. Random patterns are generated using these super-pixels. If $N$ super-pixels with a radius of $W$ are turned on, the electric field after intensity modulation is
\begin{equation}
E_A(\xi, \eta) = \sum\limits_{i=1}^N rect(\frac{\xi-\xi_i}{2W})rect(\frac{\eta-\eta_i}{2W}).\label{eq1}
\end{equation}
where $rect$ is the rectangular function describing a turned-on super-pixel. The next step is to calculate the electric field $E_0$ on the surface of scattering medium, assuming the unevenness of the surface within the beam size is negligible compared to the working distance of the objective lens. If the thickness of the scattering medium is $\Delta z_0$, $E_0$ is calculated by 
\begin{equation}
E_0(x, y) = \mathcal{P}_{-\Delta z_0}{\mathcal{F}^{-1}{E_A(\xi, \eta)}},\label{eq2}
\end{equation}
where $\mathcal{F}$ denotes Fourier transform and $\mathcal{P}_{-\Delta z_0}$ denotes the linear operator for Fresnel propagation backwards by distance $\Delta z_0$:
\begin{equation}
\mathcal{P}_{-\Delta z_0} = \mathcal{F}^{-1}{\exp[-j2\pi\Delta z_0\sqrt{1/\lambda^2-(\xi^2+\eta^2)}]\cdot\mathcal{F}}.\label{eq3}
\end{equation}
That is, the incident field on the surface of the sample is calculated by first computing the electric field at the focus, and then back-propagating it from the focus to the surface over a distance of $\Delta z_0$. By combining Eq. \ref{eq2} and Eq. \ref{eq3}, the electric field on the surface of the scattering medium can be simplified to
\begin{equation}
E_0(x, y) = \mathcal{F}^{-1} {\exp[-j2\pi \Delta z_0 \sqrt{1/\lambda^2-(\xi^2+\eta^2)}]\cdot E_A(\xi,\eta)}. \label{eq4}
\end{equation}
In matrix form, 
\begin{equation}
E_0 = F^{-1} diag(H_{\Delta z_0}^*) E_A,\label{eq5}
\end{equation}
where $H$ is the forward Fresnel propagation over distance $\Delta z_0$. The incident electric field $E_0$ propagates through the scattering medium over a distance $\Delta z_0$, eventually reaching the reference fluorescence object. The excitation field on the fluorescence object is
\begin{equation}
E_{exc} = T_{exc}E_0, \label{eq6}
\end{equation}
where $T_{exc}$ is the transmission matrix of the scattering medium at the excitation wavelength. The intensity of the excitation field $E_{exc}$ is
\begin{equation}
I_{exc} = |E_{exc}|^2\label{eq7}
\end{equation}
Since each pattern randomly selects multiple beams, they may or may not constructively interfere at a focus after propagating through the unknown scattering media. If these beams interfere constructively, the center lobe of the interference pattern will excite fluorescence through two-photon absorption if the photon density in the center lobe is high enough. Otherwise, the beams do not generate a bright focus, resulting in low or zero fluorescence intensity. Beams that create a bright focus may also be multiply scattered, as long as their phase difference is zero. In two-photon microscopy, fluorescence intensity is quadratically proportional to the intensity of excitation light in two-photon excitation, represented as
\begin{equation}
\begin{split}
I_f & = I_{exc}^2 \\
 & = |T_{exc}F^{-1} diag(H^*) E_A|^4. 
\end{split} \label{eq8}
\end{equation}
The excited fluorescence propagates backward to exit the scattering medium. Because the wavelength of the fluorescence differs from that of the excitation light, and because fluorescence is incoherent (or partial coherent) light while the excitation light is coherent, we use an intensity transmission matrix $T_f$ for fluorescence. This matrix differs from the transmission matrix $T_{exc}$ used for excitation light. The signal detected by the PMT represents the total fluorescence intensity exiting the scattering medium, and is expressed as:
\begin{equation}
\begin{split}
I & = \sum T_fI_f \\
  & = \sum T_f|T_{exc}F^{-1} diag(H^*)E_A|^4 \\
 & = \sum T_f|T_{exc}F^{-1} diag(H^*)\sum\limits_{i=1}^NE_i|^4,
\end{split} \label{eq9}
\end{equation}
where $E_i$ is the matrix form of the binary random patterns with only one super-pixel turned on. Since computing the correction mask by solving two transmission matrices from a nonlinear equation would be extremely challenging, to enable high-speed active correction, our technique probes the correction mask by simultaneously turning on multiple random pixels. During our experiments, several thousand measurements are taken in this process. Since the total fluorescence intensity is proportional to the $4^{th}$ power of the electric field of these beams, 2P-FOCUS is highly sensitive to the number of pixels that are turned on, resulting in highly effective and efficient approach for scattering correction.

The next step involves calculating the correction mask (Fig.~\ref{fig1}B, step 2). The random masks that generate bright fluorescence are selected by thresholding the fluorescence intensity detected by the PMT. Since these beams can create bright center lobes through interference, they have little phase difference. Thus, summing these random patterns generates a new pattern. The multiple beams from this new pattern are coherent and in phase, producing a brighter center lobe. This new pattern serves as the gray-scale correction mask. 

The final step is generating a binary correction mask for DMD projection by thresholding the intensity of the grayscale correction mask (Fig.~\ref{fig1}B, step 3). To ensure a fair comparison, we maintain the same illumination power on the sample before and after correction, staying below the thermal damage threshold of biological tissues. This is a unique aspect of our method compared to conventional intensity modulation. For example, if a blank mask is used before correction, allowing no light blockage (Fig.~\ref{fig1}A), we then increase the input power to the DMD when applying correction masks to compensate the power blocked by turned-off pixels (Fig.~\ref{fig1}C). Power control can also be achieved by using different thresholds to binarize the grayscale correction mask. Different thresholds result in varying illumination power on the sample, given the same input power to the DMD, which will be discussed in detail in the following sections. Since all beams selected by the grayscale correction mask are coherent and in phase, the subset of these beams after binarization remains coherent and in phase. Their interference produces a brighter center lobe compared to the scenario without scattering correction, not only effectively redistributing part of the illumination power from out-of-phase beams to in-phase beams but also nonlinearly increasing the laser power in the center lobe due to coherent interference.

\subsection{Experimental Setup and Calibration}\label{subsec22}
The experimental setup is depicted in Fig.~\ref{fig1}D and described in Appendix~\ref{setup}. We utilize a femtosecond laser with a 1035 nm wavelength for two-photon excitation. Our setup includes a near-infrared coated DMD with 1280 $\times$ 800 pixels, of which only the central 800 $\times$ 800 pixels were used, and a maximum patterning rate of 12.5 kHz. In the following experiments, the DMD is operated at a 1 kHz patterning rate due to limitations of other electronics. Since the DMD is located on the conjugate plane of the back aperture, it induces spatial dispersion in the femtosecond pulses, resulting in an elliptical focus (Fig.~\ref{fig1}E, left). To correct this spatial dispersion and compress the pulses, we implement a reflective diffraction grating on the conjugate plane of the DMD \cite{Geng2019-rt, Wang2020-oy}. The parameters of the grating, the 4-$f$ relay lenses (L3, L4), and the incident angles are carefully designed and aligned to fully compensate for the spatial dispersion. It is important to note that the effective groove size of the DMD differs from the micromirror pitch size. After dispersion compensation, the point-spread-function (PSF) becomes a tight circular spot (Fig.~\ref{fig1}E, right). A two-axis scanning mirror system, operating at 1 kHz (the maximum speed of the galvo-scanner), is also positioned on the conjugate plane of the DMD. This system scans the focus across the field-of-view and currently represents the speed bottleneck of the system. The scanning mirrors are synchronized with the DMD, allowing the application of different correction patterns to specific subregions as necessary. The modulated light field is relayed to the back aperture via 4-$f$ systems and then focused by the objective lens onto the scattering sample. The sample is placed on a linear translation stage for capturing $z$-stack images. Two-photon excited fluorescence is collected in reflection mode by the objective lens and detected by a PMT. The 3D PSF of 2P-FOCUS is 1.7 $\times$ 1.7 $\times$ 6.6 $\mu m^3$, measured with 0.71-$\mu m$ red fluorescent beads without scattering media (Fig.~\ref{fig1}F-G). The resolution is slightly lower than theoretical values because the surface of the DMD induces astigmatism, which could potentially be corrected by adding a SLM to the system.

\section{Results}\label{sec3}
\subsection{Focusing through Bone with 2P-FOCUS}\label{subsec31}
\begin{figure}[h!]%
\centering
\includegraphics[width=0.82\textwidth]{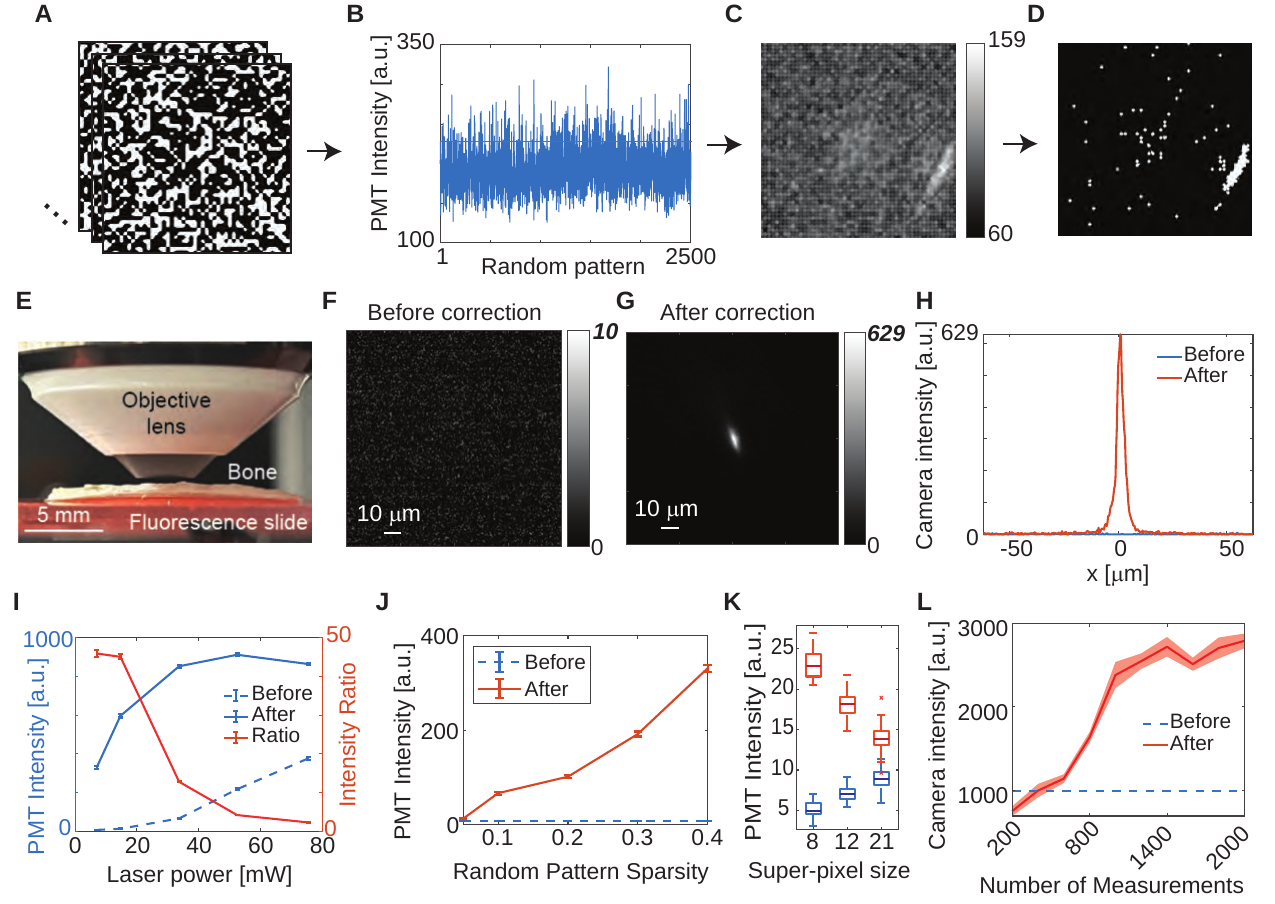}
\caption{Focusing through bone with 2P-FOCUS. (A) Representative random patterns with a sparsity of 0.4 and a super-pixel radius of 8 pixels used in the experiment. (B) Fluorescence intensity corresponding to the 2,500 patterns detected by the PMT in reflection mode. Random patterns contributing to the top 10\% of fluorescence intensity are selected (above the red line). (C) The sum of the selected random binary patterns forms the grayscale correction mask. (D) The final correction mask is generated by binarizing the grayscale correction mask. (E) Photo of the bone sample and the fluorescence sample. (F-G) Zoomed-in fluorescent image of the focus (F) before correction and (G) after correction taken with a camera in reflection mode. (H) Comparison of the intensity profile along the $x$-axis before (blue line) and after (red line) correction. (I) Fluorescence intensity before (blue dashed line) and after (blue solid line) correction, as well as the intensity ratio (red line), as functions of the laser power on the sample. (J) Fluorescence intensity before (blue) and after (red) correction as a function of the sparsity of random patterns with a super-pixel radius of 8 pixels. (K) Fluorescence intensity before (blue) and after (red) correction as a function of the size of super-pixels. Fluorescence intensity before correction is measured with random patterns in this plot. (L) Fluorescence intensity before (blue) and after (red) correction as a function of the number of measurements.}\label{fig2}
\end{figure}

We first demonstrate 2P-FOCUS experimentally by focusing through a chicken bone onto a homogeneous fluorescence slide (Fig.~\ref{fig2}). We adhered a piece of chicken bone without any exogenous fluorescent labels onto a microscope slide coated with a thin layer of red fluorescent paint (Fig.~\ref{fig2}E). The surface of the bone is uneven and the part we focused through is about 200 $\mu$m thick. The bone is highly scattering and no focus is visible before correction at an illumination power of 7 mW (Fig.~\ref{fig2}F). The mean and standard deviation of intensity of Fig.~\ref{fig2}F is 0.58$\pm$1.01. The maximum value of Fig.~\ref{fig2}F comes from a hot pixel. The random patterns used in the experiment has a sparsity of 0.4 and a super-pixel radius of 8 pixels (see next paragraph for details). During the correction process, we took measurements with 2,500 random patterns (Fig.~\ref{fig2}A) at an illumination power of 76 mW (the power is only used for taking measurements under random patterns) and recorded the corresponding fluorescence intensity (Fig.~\ref{fig2}B). The random patterns exhibit broad coverage of spatial frequencies, with consistent frequency coverage across all patterns. They maintain the same power in the DC component to ensure high signal-to-noise ratio measurements at the same location, while the power in the high-frequency components varies slightly to probe the spatial components that propagate through the scattering bone (Appendix Fig.~\ref{figS1}A-B). We then selected the random patterns corresponding to the top 10\% brightest fluorescence (above the red line, Fig.~\ref{fig2}B, 250 patterns) and summed them to form the grayscale correction mask (Fig.~\ref{fig2}C). The grayscale value of the correction mask indicates the number of patterns that turn the super-pixel on. For instance, the peak value of 159 in Fig.~\ref{fig2}C, located on the right bottom side in the Fourier domain, shows that most random patterns (159 out of 250) turned this pixel on to create a bright focus. We then adjusted the half-wave-plate (HWP) to maximize the power input to the DMD, and binarized the grayscale correction mask (Fig.~\ref{fig2}D) using a threshold that ensures the illumination power on the sample remains at 7 mW after modulation by the binary correction mask, which is the same as before correction. In this case, the correction mask only turns on pixels that can generate highly constructive interference (see next paragraph for details). Given the turned-on pixels are mostly on the edge of the back aperture in this case, our interpretation is that the bone is probably uneven in the field-of-view and the correction mask tends to guide light through the relatively thinner (less scattering) part. The correction of a relatively flat bone is shown in Fig.~\ref{fig3}. This correction mask leads to a bright fluorescent focus with a peak intensity of 629 (Fig.~\ref{fig2}G), measured with a camera in the \textit{reflection mode} (replacing the PMT in Fig.~\ref{fig1}D) for visualization purposes only. The distorted shape of the fluorescence spot is due to the scattering of fluorescence as it propagates back through the bone, as the correction applies only to the excitation beam and not the emitted fluorescence. This is not an issue in two-photon microscopy, as scattered emission fluorescence can still be collected by the PMT as long as it falls within the numerical aperture (NA) of the objective lens. The intensity profile across the $x$-axis (Fig.~\ref{fig2}H) also demonstrates the significant improvement in fluorescence intensity before and after correction.

We next explore how various factors influence the correction result, including the thresholds for binarization, the sparsity of random patterns, the radius of super-pixels, and the number of measurements. The binarization from Fig.~\ref{fig2}C to Fig.~\ref{fig2}D depends on the threshold applied to the grayscale correction mask, which controls the illumination power applied to the sample given the same input power to the DMD. The experiment depicted in Fig.~\ref{fig2}A-H describes a scenario where the laser power on the sample is 7 mW, corresponding to the first data point in Fig.~\ref{fig2}I. The fluorescence intensity value shown in Fig.~\ref{fig2}I is measured by the PMT, so it differ from the absolute values in Fig.~\ref{fig2}F-H. The ``before” group in Fig.~\ref{fig2}I is measured under random patterns, and the power for the ``before” group is adjusted by tuning the HWP. Random patterns instead of a blank screen are used here to check whether the random pattern itself causes a non-quadratic relationship between the illumination power and the fluorescence intensity.  As the illumination power increases (to 15 mW, 34 mW, 52 mW, and 76 mW), the fluorescence intensity before correction increases quadratically with the illumination power (blue dashed line in Fig.~\ref{fig2}I, and Appendix Fig.~\ref{figS1}C), consistent with the principles of two-photon excitation. Importantly, after applying the correction mask, the intensity of two-photon excited fluorescence \textit{does not follow the quadratic rule} as the excitation power increases by turning on more pixels on the DMD (blue solid line in Fig.~\ref{fig2}I, and the corresponding correction masks are shown in Appendix Fig.~\ref{figS1}D-E). This means the significance of each pixel on the DMD is different. The brightest pixels in Fig.~\ref{fig2}C generate ``most in-phase beams", while other bright pixels generate ``partially in-phase beams". Turning on the DMD pixels corresponding to the partially in-phase beams increases the illumination power on the scattering sample, but only part of the laser power is delivered to the fluorescence object and constructively interferes with other beams. In addition to partially in-phase beams, several other factors can also increase the power on the sample without contributing to two-photon fluorescence. One such factor is the activation of pixels corresponding to low-transmission modes, as each pixel on the DMD corresponds to different spatial modes with varying transmission through the tissue (Fig.~\ref{fig2}C). These beams may also be diffracted away from the focus of constructive interference. The most significant improvement in fluorescence intensity is achieved when the laser power on the sample is relatively low. In other words, the strategy to achieve a large improvement after correction is making the correction mask turn on only the pixels generating highly in-phase beams after scattering (red line, Fig.~\ref{fig2}I). This result clearly proves the mechanism of 2P-FOCUS is selecting the multiple beams that are still highly in phase after scattering while blocking those out-of-phase ones. 

The second factor is the sparsity of random patterns, which indicates the percentage of pixels that are turned on out of the total number. For random patterns with a super-pixel radius of 8 pixels, there are $50 \times 50 =$ 2,500 super pixels in each mask, and a total of 2,500 masks are used, forming a 2500 $\times$ 2500 matrix $P$. The sparsity of random patterns ranges from 0.05 to 0.9 (see examples in the Appendix Fig.~\ref{figS2}A). To demonstrate that the random patterns are orthogonal to each other, we calculated $I = P^TP$ using normalized random patterns with a sparsity of 0.4. The histogram of all elements in $I$ shows that all diagonal diagonal values equal 1 and non-diagonal values are $0\pm0.02$, indicating that $I$ is approximately an identity matrix    (Fig.~\ref{figS2}B). Consequently, the random patterns are orthogonal to each other. The sparsity of random patterns is also related to illumination power. With the same input laser power to the DMD, less sparse patterns block less input power and emit more power to the sample. This increases the SNR of the measurements, consequently improving the accuracy of the correction masks. In the case shown in Fig.~\ref{fig2}J (the values are listed in Table~\ref{TableS1}), we first measure the fluorescence intensity using random patterns at 7 mW, as in the pre-correction case, and then take measurements under different sparsity (0.05, 0.1, 0.2, 0.3, and 0.4), while keeping the input power to the DMD constant. These measurements generate five different correction masks, each with an output power of 7 mW (see Appendix Fig.~\ref{figS2}C). We also took measurements under sparsity ranging from 0.2 to 0.5 through another region of the bone and found that the fluorescence signal decreases at the sparsity of 0.5 (see Appendix Fig.~\ref{figS2}D-E and Table~\ref{TableS2}). The correction mask derived from the random patterns with 0.4 sparsity shows the most significant improvement in fluorescence intensity in both cases, as depicted in Fig.~\ref{fig2}J and Fig.~\ref{figS2}D. In practice, the optimal sparsity of random patterns varies depending on the scattering properties of the sample, and the brightness of the fluorescence signal, and the maximum power of the laser. 

The third factor influencing the correction result is the size of super-pixels. The center square region of the DMD consists of $800 \times 800 = 640,000$ pixels. Well-sampling all these pixels requires 640,000 random patterns, which is too large for the on-chip RAM of the DMD to handle and also requires more computing and experimental time. Therefore, we binned these pixels into super-pixels to accelerate the process. For super-pixels with a radius of $W$, the number of random patterns to well-sample the frequency domain is $800^2/(2W+1)^2$. We compared the results generated by patterns with $W$ = 8, 12, and 21 (see Fig.~\ref{fig2}K). The fluorescence intensity before correction was measured under random pattern modulation with a sparsity of 0.05. The illumination power on the sample was approximately 12 mW, similar before and after correction for all three pattern types. Before correction, fluorescence intensity increases as the super-pixel size increases due to the effect of diffraction. Focusing a beam modulated by random patterns on the back aperture generates a main lobe of focus with speckles due to the diffraction of the random patterns. Random patterns with smaller super-pixel sizes diffract more than those with larger super-pixel sizes, resulting in lower peak power at the main lobe and higher power at the speckles given the same total input power. After correction, correction masks with smaller super-pixels correct scattering more effectively because they identify in-phase beams with greater precision. 

We also explored how the number of measurements influences scattering correction. For random patterns with a super-pixel radius of 12 pixels, the minimum number of measurements required for well-sampling is 1,024. We generated correction masks at 0.01 sparsity with various numbers of raw measurements, ranging from 200 to 2000, and compared the results of the scattering correction. As shown in Fig.~\ref{fig2}L, when the number of measurements is too small (i.e., 200 measurements in this case) to reconstruct an accurate correction mask, this failed mask generates a main lobe of focus and speckles on the image plane, similar to a random pattern. This leads to the fluorescence excited by the focus being even lower than that without correction, which projects a blank mask on the back aperture like standard two-photon microscopy. The blank mask is used to measure the baseline, as it is commonly used in standard two-photon microscopy. The fluorescence intensity improves with an increasing number of measurements, but this improvement becomes marginal when the measurements become redundant. Considering the computing and experimental time, the results suggest that the optimal number of measurements corresponds to the minimum needed to adequately sample the Fourier domain.

\subsection{Subregion correction for imaging beyond the memory effect range}\label{subsec32}
\begin{figure}[htp!]%
\centering
\includegraphics[width=0.82\textwidth]{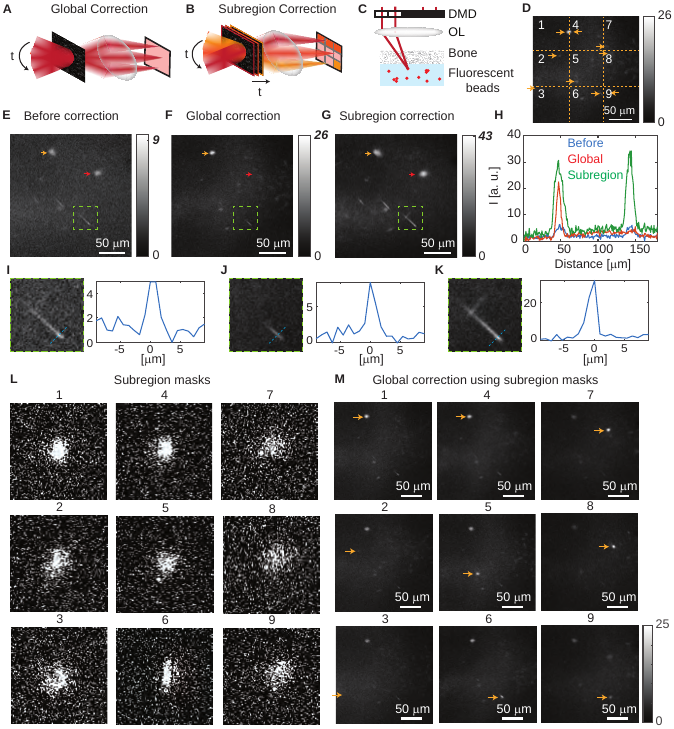}
\caption{Imaging fluorescence beads through a 200 $\mu$m-thick bone beyond the memory effect range. (A) Schematic diagram of global correction. Global correction applies the same correction mask to all scanning locations during the imaging process. (B) Schematic diagram of subregion correction. Subregion correction applies a different correction mask for each subregion. The scanning mirror and the DMD are synchronized to project the corresponding mask at each scanning location. (C) Schematic diagram of the sample used in this experiment. A piece of chicken bone was adhered on top of red fluorescent beads in PDMS. (D) The $3 \times 3$ subregions and corresponding references. (E-H) Imaging fluorescent beads through highly scattering bone across a 230$\times$230 $\mu m^2$ field-of-view. The laser power on the sample is 25 mW for all three cases. (E) The image taken without scattering correction. (F) The image taken under global correction. The reference object is indicated by the yellow arrow. The object outside the memory effect range is invisible (indicated by the red arrow). The peak intensity improved from 9 to 26 after global correction. (G) The image taken under subregion correction with the reference objects and correction masks in (D) and (I). With subregion correction, the previously invisible object in (F) becomes visible (pointed by the red arrow). The peak fluorescence intensity improved from 9 to 43 after subregion correction. (H) The intensity profile of the two clusters of beads indicated by arrows in (E-G). The distance between these two objects is 96 $\mu$m. The correction mask for imaging the left object cannot effectively correct scattering for imaging the right object, indicating the memory effect range of this region is smaller than 96 $\mu$m. (I-K) Zoomed-in views of the bead in the green box in (E-G) and their intensity profiles across the green line. (L) The correction masks for the 9 subregions generated using the references pointed out in (D). (M) Images taken by applying each subregion mask as a global mask. These images clearly show different masks correct scattering differently even though they look similar to each other. The images are displayed under the same color bar for comparison.}\label{fig3}
\end{figure}

We next explore the optimal strategy for imaging a large field-of-view through highly scattering media using 2P-FOCUS. 2P-FOCUS can correct scattering during the imaging process with point scanning; it scans the focus across the field-of-view pixel by pixel while applying correction masks. The most straightforward strategy is to apply the same correction mask for all scanning positions, referred to as ``global correction" (Fig.~\ref{fig3}A). Global correction is valid if the scanning area is within the memory effect range \cite{Freund1988-va}. However, for severe scattering media, the scattering properties are highly spatially varying, resulting in a limited memory effect range. The range is typically a few tens of microns for highly scattering biological tissue \cite{Yoon2020-sm}, depending on the thickness of the scattering layer and the wavelength of the incident light. In this case, the field-of-view for imaging through scattering tissue is limited by the memory effect range.

To image beyond the memory effect range, we developed a new strategy called ``subregion correction" (Fig.~\ref{fig3}B), which synchronizes the scanner and the DMD to seamlessly project different masks to different regions during the scanning process. To ensure the speed of correction, we developed control code that can automatically perform subregion correction. First, it captures the image before correction, splits the whole image into multiple subregions (e.g., $3\times3$), and automatically selects the pixel with the maximum intensity as the reference for each subregion, which can be fluorescence-labeled objects or background fluorescence. If the fluorescence intensity of a reference is too low, indicating it may be noise, the subregion is merged into the adjacent one. Next, the code automatically directs the scanner to each reference and generates the corresponding correction mask. The same set of random patterns is used in every subregion and preloaded into the DMD's RAM to minimize data communication time. After generating all subregion correction masks, the code synchronizes the scanner and the DMD to apply each subregion-correction mask to the corresponding region while the sample remains steady. The scanner operates at its maximum speed of 1 kHz during the process. This seamless and automated control significantly accelerates the subregion correction process, making scattering correction beyond the memory effect range feasible in practice.

We experimentally compared global correction and subregion correction side-by-side by imaging red fluorescent beads through a piece of chicken bone (Fig.~\ref{fig3}C). The size of the fluorescent beads is 0.71 $\mu$m, suspending in PDMS, with some clusters. The bone we imaged through is about 200 $\mu$m thick. The field-of-view of imaging is 230$\times$230 $\mu m^2$, and the post-objective power is 25 mW for all image conditions. We first imaged the sample without any correction by projecting a blank mask at all scanning locations. The image (Fig.~\ref{fig3}E) shows a low signal-to-noise ratio, revealing two bead clusters and several single beads with severe aberrations. Because the bone on top is highly heterogeneous, the aberration and scattering caused by the bone are also spatially varying. We next performed a global correction using the bead cluster indicated by the yellow arrow in Fig.~\ref{fig3}E. The correction used 10,000 random patterns with a super-pixel size of 4 pixels and a sparsity of 0.4. By applying the same correction mask to all scanning locations, the fluorescence intensity after global correction improved by 2.9 times (Fig.~\ref{fig3}F). However, only the fluorescent bead cluster in the area where the correction mask was measured becomes brighter (indicated by the yellow arrow in Fig.~\ref{fig3}E-F), while the fluorescent bead cluster that is 96 $\mu$m away (indicated by the red arrow in Fig.~\ref{fig3}E-F) did not. 

Next, we conduct subregion correction on the same field-of-view. The image taken under global correction is divided into $3\times3$ subregions (Fig.~\ref{fig3}D), and the peak intensity of each subregion is automatically selected as the reference for that subregion (pointed by yellow arrows in Fig.~\ref{fig3}D). Both fluorescent beads and background fluorescence can serve as references to generate correction masks, with the selection being automatically made by the code based on intensity. Brighter objects provide higher signal-to-noise ratio measurements, resulting in more accurate correction masks. The same set of random patterns used in global correction is applied to generate the correction mask for each subregion. Under subregion correction, the image (Fig.~\ref{fig3}G) shows improved intensity and enhanced contrast across the entire 230$\times$230 $\mu m^2$ field-of-view. Objects that were previously invisible under global correction become visible (indicated by the red arrow in Fig.~\ref{fig3}E-G). The peak intensity improved by 4.8 times compared to the pre-correction case. The intensity profile across the two bead clusters quantitatively compares the results of global and subregion correction (Fig.~\ref{fig3}H). The line plot indicates that the memory effect range of this bone sample is less than 96 $\mu$m, while the thickness of the bone is about 200 $\mu$m. Therefore our method corrects multiple scattering. Additionally, to examine potential photobleaching due to the correction process, we imaged the same region with a blank mask before and after the correction process (Appendix Fig.~\ref{figS3}). The result shows the peak intensity slightly decreased from 9 to 8, indicating only minor photobleaching. Both global and subregion correction effectively improve the fluorescence intensity of objects within the memory effect range (left peak in Fig.~\ref{fig3}H), while only subregion correction improves intensity beyond the memory effect range (right peak in Fig.~\ref{fig3}H). 

We next check the image resolution using a single bead before correction, after global correction, and after subregion correction (Fig.~\ref{fig3}I-K). Before scattering correction, the image of a single bead is elongated due to strong aberrations from the bone, with a FWHM of 2.48 $\mu$m along the short axis and an average peak intensity of 4.9 (Fig.~\ref{fig3}I). After global scattering correction, where the reference is outside of the memory effect range, the bead still looks elongated with a FWHM of 1.91 $\mu$m and an average peak intensity of 8.23 (Fig.~\ref{fig3}J). The slight increase in fluorescence intensity is probably because the scattering potential in this region is coincidentally similar to the region where the correction is conducted. After subregion correction, the bead is still elongated with a FWHM of 1.87 $\mu$m, but the peak intensity improves to 31.55 (Fig.~\ref{fig3}K). Unlike phase modulation or intensity modulation for aberration correction \cite{Ren2020-xy, Chen2020-qv}, our method selects in-phase beams to maximize fluorescent intensity passing through scattering media without correcting aberrations. Our technique could be further enhanced when combined with aberration correction methods, which is out of the scope of this paper. The results quantitatively demonstrate the intensity modulation does not reduce spatial resolution compared to the case before correction when all pixels on the DMD are turned on, because the blocked beams cannot pass through the scattering sample and constructively interfere even if the corresponding DMD pixels are turned on.

To further demonstrate our approach for correcting spatially varying scattering, we use subregion masks generated with different references (Fig.~\ref{fig3}D) as global correction masks to scan across the entire field-of-view. The ``on" pixels in the correction patterns (Fig.~\ref{fig3}L) are predominantly located near the center, likely due to the Gaussian beam profile and global aberration. Assuming the surface of the sample is relatively flat, on-axis ballistic photons propagate a shorter distance than off-axis ballistic photons, resulting in more on-axis ballistic photons reaching the focus. On the other hand, the correction masks function differently in correcting scattering at various locations (Fig.~\ref{fig3}M) because of their high-frequency features. For example, mask \#1 and mask \#4, calculated using the left bead cluster (pointed by the yellow arrow in Fig.~\ref{fig3}E-G), brighten the left cluster but not the right, while mask \#7 and mask \#8, calculated with the right bead cluster (pointed by the red arrow in Fig.~\ref{fig3}E-G), do the opposite. Mask calculated from relatively dim references, like \#2, \#3 and \#9, show less effective correction. The results indicate that 2P-FOCUS improves fluorescence intensity by scattering correction, since the mask tailored to one specific scattering regions cannot correct scattering in another region beyond the memory effect range. Subregion correction is necessary to address scattering outside the memory effect range across a large field-of-view.

To further explore the contribution of high-frequency and low-frequency components in the correction masks, we applied various smoothing filters to the correction masks to partially remove the high-frequency components (Fig.~\ref{figS3}A, plot 1-5). Additionally, we applied a low-NA mask as a limiting case that excludes all high-frequency components (Fig.~\ref{figS3}A, plot 6), which has been used to enhance imaging depth in two-photon microscopy \cite{Tung2004-ew, Kondo2017-lh}. The illumination power on the sample was kept constant under all masks. The corresponding fluorescence intensity under each mask (Fig.~\ref{figS3}B-C) demonstrates that maximum fluorescence intensity is achieved after slight smoothing (plot 4). In contrast, both the absence of smoothing and excessive smoothing (plot 5-6) result in a drop in fluorescence intensity. Without smoothing, the high-frequency components diffract light severely, reducing the power in the central lobe. On the other hand, excessive smoothing or the low-NA mask fails to effectively correct highly spatially varying scattering, also reducing the power of the focus. Plot 3 and plot 4 exhibit similar intensities, indicating that the positions of some super-pixels are less significant than others. This observation aligns with our findings in Section 3.1, where we noted that beams from certain super-pixels can constructively interfere while others cannot. These results demonstrate that 2P-FOCUS primarily improves fluorescence intensity by correcting highly spatially varying scattering, with aberration correction playing a secondary role, and fundamentally differs from approaches that enhance imaging depth using a low NA.

\subsection{Imaging neurons deep in the brain with global and subregion scattering correction.}\label{subsec33}
\begin{figure}[h!]%
\centering
\includegraphics[width=0.82\textwidth]{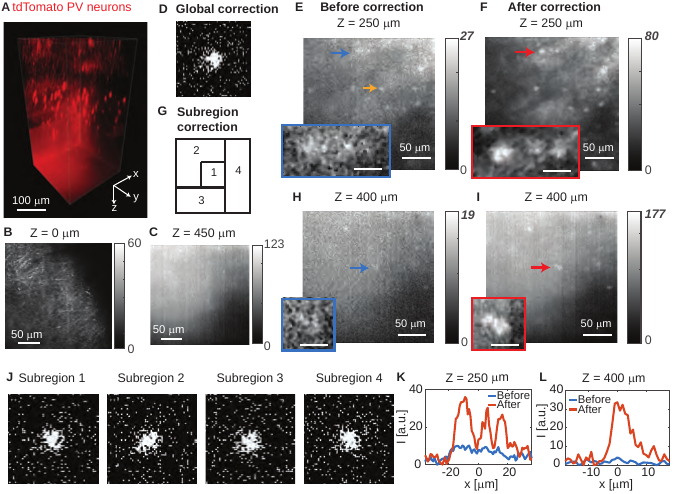}
\caption{Imaging fluorescence-labeled neurons deep in the mouse brain ex vivo using 2P-FOCUS. (A) The volumetric view of parvalbumin (PV) interneurons expressing cell-fill tdTomato imaged by 2P-FOCUS. The image volume is 230$\times$230$\times$450 $\mu m^3$. (B) The top plane and (C) the bottom plane of the image stack. The image stack is shown in Supplementary Visualization 1. (D) The correction mask for global correction at 250 $\mu$m depth, generated with the neuron pointed by the yellow arrow in (E). (E-F) Images of the same region at 250 $\mu$m depth (E) before and (F) after global correction. Fluorescence intensity is improved by about 3-fold after correction. Note that these two figures are displayed with different color bars to visualize the dim objects before correction. The same figures displayed with the same color bar are shown in Appendix Fig.~\ref{figS3}D. Insert: Zoomed-in view of the neurons pointed by the blue and red arrows. Scale bar, 5 $\mu$m. (G) The four subregions. (H-I) Images of the same region at 400 $\mu$m depth (H) before and (I) after subregion correction. Fluorescence intensity is improved by about 9.3-fold after correction. Insert: Zoomed-in view of the neurons pointed by the blue and red arrows. Scale bar, 5 $\mu$m. (J) The four subregion correction masks. (K) Comparison of the intensity profile of representative neurons (zoomed-in view in (E-F)) at 250 $\mu$m depth before (blue) and after (red) correction. (L) Comparison of the intensity of representative neurons (zoomed-in view in (H-I)) at 400 $\mu$m depth before (blue) and after correction (red).}\label{fig4}
\end{figure}

We next demonstrate the imaging capabilities of 2P-FOCUS by imaging fluorescence-labeled neurons in the mouse brain. In our first imaging experiment, we imaged a 230$\times$230$\times$450 $\mu m^3$ volume in the primary visual cortex of an intact transgenic mouse brain with cell-fill expression of tdTomato in all parvalbumin (PV) interneurons (see Fig.~\ref{fig4}A). The mouse underwent transcardial perfusion with 4\% Paraformaldehyde (PFA) and was then preserved in 4\% PFA until imaging (see Appendix~\ref{sample} for detailed sample preparation). During the imaging session, the whole brain was maintained in phosphate-buffered saline and placed beneath the objective lens. The brain was imaged from the top downward. The top plane ($Z=0$) was not on the brain's surface but below it, as dendrites and axons are observed on the plane (Fig.~\ref{fig4}B). When imaging the top 250 $\mu$m, we did not apply any correction but gradually increased the illumination power on the sample from 2 mW to 14 mW. The first correction was applied when imaging the plane at a 250 $\mu$m depth (see Fig.~\ref{fig4}D-F). Before correction, the image contrast is low, but we can still identify some bright neurons (Fig.~\ref{fig4}E). We selected one of the bright neurons near the center of the field-of-view, indicated by the yellow arrow in Fig.~\ref{fig4}E, as the reference object. It is important to note that the reference neuron is not a guide star, as it does not need to be a point source; any fluorescent object can be used as a reference. We applied an offset voltage to the scanning mirrors to focus on the reference neuron, and then performed a scattering correction to maximize the fluorescence intensity at this neuron. We used 2,500 random masks with a super-pixel radius of 8 pixels and a sparsity of 0.4 for measurements, and generated a correction mask as shown in the bottom of Fig.~\ref{fig4}D. We applied this correction mask to all scanning positions as a global correction, and acquired the image of the same region after correction (Fig.~\ref{fig4}F). The maximum fluorescence intensity is improved by about 3-fold (from 27 to 80). The intensity of the reference neuron is improved by about 1.8-fold (see Appendix Fig.~\ref{figS4}A-C), which is less than the maximum improvement and may indicate minor photobleaching during the correction process. The image contrast is also improved after correction, as the three neurons in the zoomed-in view (Fig.~\ref{fig4}E-F) were not distinguishable before correction but became distinguishable after correction. The intensity profile of the three neurons are plotted in Fig.~\ref{fig4}K.

As the imaging depth increases, the global correction mask no longer provides high intensity and high contrast images when the imaging region is larger than the range of the memory effect. We applied subregion correction to image deep in the mouse brain. We first conduct subregion correction at 290 $\mu$m where the peak intensity drops to the half of the peak intensity of the image at 250 $\mu$m depth after global correction (see Appendix Fig.~\ref{figS4}D-F), indicating the image region at 290 $\mu$m depth is out of the memory effect range from the image region at 250 $\mu$m. The field-of-view is divided into 3 subregions, and three correction masks are generated using reference fluorescence at 290 $\mu$m (see Appendix Fig.~\ref{figS4}G). Notice that one of the reference fluorescence is the background fluorescence rather than a neuron. The three correction masks are applied when imaging the next few $z$ planes ($z = 290~\mu m - 380~\mu m$). The peak intensity of the image at 380 $\mu$m using the correction masks generated at $z = 290~\mu m$ is 66 (see Appendix Fig.~\ref{figS4}4J), indicating this imaging area was out of the memory effect range from the imaging area at 290 $\mu$m depth. Thus, we conduct another round of subregion correction at 380 $\mu$m depth and generated four correction masks (Appendix Fig.~\ref{figS4}K-L, Fig.~\ref{fig4}J). These masks are then applied for imaging $z$ planes at $380~\mu m - 450~\mu m$ depth. Fig.~\ref{fig4}H-I show the image at 400 $\mu$m depth before and after subregion correction. The subregion correction successfully improved the fluorescence intensity by 9.3 times (see Fig.~\ref{fig4}I) compared to the image without correction (Fig.~\ref{fig4}H). The neuron barely visible before correction becomes visible, indicating an improvement in the SNR (see zoomed-in views in Fig.~\ref{fig4}H-I). The intensity profile of the representative neuron pointed by the arrows is shown in Fig.~\ref{fig4}L. The improvement is less significant than that achieved when focusing through bone, because the correction mask used for focusing through bone is more selective than the one used for imaging neurons. The fluorescence slide used in the focusing experiment is much brighter than the fluorescence-labeled neurons in the mouse brain. Given the same input laser power, with brighter fluorophores, the correction mask can selectively turn on only the pixels corresponding to highly in-phase beams and still achieve a sufficient SNR. In contrast, with less bright fluorophores, the correction mask has to turn on more pixels, including those corresponding to partially in-phase beams. Therefore, the intensity improvement in brain imaging is less than that in focusing through bone, which is consistent with our result in Fig.~\ref{fig2}I. The bottleneck is the power density of the input laser to the DMD. Subregion correction enables us to image a 230$\times$230 $\mu m^2$ region deep within the scattering brain tissue, achieving better results than global correction.

\subsection{Imaging cerebral blood vessels deep in the mouse brain by precise modulation.}\label{subsec34}
\begin{figure}[h!]%
\centering
\includegraphics[width=0.82\textwidth]{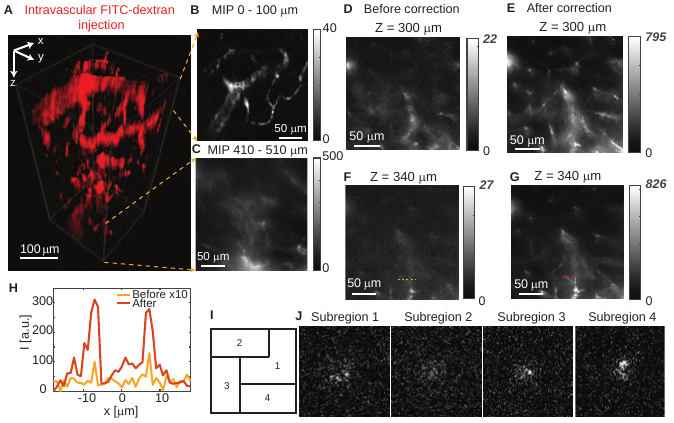}
\caption{Imaging blood vessels with intravascular fluorophore injection deep in the mouse brain using 2P-FOCUS. (A) Volumetric view of cerebral blood vessels with intravascular FITC-dextran injection, imaged by 2P-FOCUS ex vivo. The image volume is 230$\times$230$\times$510 $\mu m^3$. (B) Maximum intensity projection (MIP) of the top 100 $\mu$m-thick volume along the $z$ axis. (C) MIP of the bottom 100 $\mu$m-thick volume along the $z$ axis. The image stack is shown in Supplementary Visualization 2. (D-E) Two-photon image of blood vessels at 300 $\mu$m depth (D) before and (E) after correction. The peak intensity is improved from 22 to 795, corresponding to 36.1-fold improvement. (F-G) Two-photon image of blood vessels at 340 $\mu$m depth (F) before and (G) after correction. All pixels of the DMD are turned on when capturing (F). The peak fluorescence intensity is increased from 27 to 826, corresponding to 30.6-fold improvement. (H) Comparison of the fluorescence intensity profile before (yellow line) and after (red line) correction along the dashed line in (F-G). The intensity profile of the case before correction is magnified by 10 times for display purposes. (I) The location of the 4 subregions on the image plane. (J) The correction masks used in the experiment when acquiring image (E) and (G).}\label{fig5}
\end{figure}

To further improve the correction and demonstrate the broad applicability of 2P-FOCUS, we use finer random masks with a super-pixel radius of 4 pixels at 0.4 sparsity and imaged another mouse brain ex vivo, in which cerebral blood vessels are labeled via intravascular FITC-dextran injection (Fig.~\ref{fig5}A). Similar to the setup in Fig.~\ref{fig4}, the whole mouse brain was immersed in phosphate-buffered saline and imaged from the top downward. The top plane was not on the brain's surface but below it, as indicated by the visible cross-sections of cerebral blood vessels. The top 270 $\mu$m volume was imaged without correction, meaning all pixels of the DMD are turned on. Since blood vessels span three-dimensionally, we choose to show the maximum intensity projection rather than a single plane. The maximum intensity projection of the top 100 $\mu$m-thick volume is shown in Fig.~\ref{fig5}B, where the cerebral blood vessels are clearly visible. As the imaging depth increased, the laser power on the sample is gradually increased from 3 mW (for $z = 0 - 160~\mu m$) to 19 mW (for $z = 250 - 510~\mu m$). The first scattering correction is conducted at a depth of 270 $\mu$m in three subregions, using 2,500 random patterns with a super-pixel radius of 8 pixels in each subregion. At a depth of 300 $\mu$m, we conducted another subregion correction with four subregions, using 8,100 random patterns with a super-pixel radius of 4 pixels in each subregion. Compared to the image before correction (Fig.~\ref{fig5}D), the image after correction (Fig.~\ref{fig5}E) reveals fine structures of cerebral capillaries that are invisible before correction. The peak intensity is improved by a factor of 36.1 after correction, which is more than the state-of-the-art techniques for correcting scattering in the mouse brain tissue (Appendix Table \hyperlink{appendixtable}{\ref*{table1}}). The four correction masks and subregions are shown in Fig.~\ref{fig5}I-J. The same masks were used when imaging at a depth of 340 $\mu$m (Fig.~\ref{fig5}F-G), Compared to the uncorrected image with the same illumination power (Fig.~\ref{fig5}F), where all pixels on the DMD were turned on, these capillaries were barely visible. The fluorescence intensity of a cross-section of representative capillaries improves by a factor of 30.6 after correction (Fig.~\ref{fig5}H). Another example for scattering correction at 390 $\mu$m depth with six subregions is shown in the Appendix Fig.~\ref{figS5}A-C). At the bottom of the volume (410 $\mu$m --- 510 $\mu$m depth), blood vessels remain visible after scattering correction (Fig.~\ref{fig5}C). Image resolution shows no obvious change after applying the correction mask, as fine structures like capillaries with a diameter of around 2 $\mu$m are resolvable (Appendix Fig.~\ref{figS5}D-H). Compared to the results in Fig.~\ref{fig4}, the outcomes in Fig.~\ref{fig5} show better scattering correction, attributable to the use of patterns with smaller super-pixels that generate more precise correction masks. Additionally, the fluorescence labeling of blood vessels is brighter than that of the neurons, despite 1035 nm not being the optimal two-photon excitation wavelength for FITC-dextran. Furthermore, in the experiment, the mouse brain is not fixed by 4\% PFA, which results in less autofluorescence compared to the fixed mouse brain shown in Fig.~\ref{fig4}. Therefore, the SNR of measurements under random pattern modulation is higher, resulting in a more accurate correction mask.

\section{Discussion}\label{sec4}
We have demonstrated that 2P-FOCUS is a novel and fast method for correcting scattering in two-photon microscopy. The key advantages of 2P-FOCUS are its speed and effectiveness, achieved through the innovative joint design of hardware and software. In terms of hardware, 2P-FOCUS uses a DMD, which has a much faster patterning rate than an SLM, to select in-phase beams, enabling high-speed patterning at a lower cost. In terms of software, 2P-FOCUS utilizes random patterns as an orthogonal basis and a single-shot algorithm to complete the entire correction process within a few seconds depending on the number of measurements. For example, collecting 2,500 measurements at a 1 kHz sampling rate takes 2.5 seconds, calculating the correction mask takes 0.5 seconds, and projecting the correction mask takes 1 ms---a total of 3 seconds to complete the entire correction process, with no iterations needed. In each subregion correction task, the random patterns used for all subregions are identical and preloaded into the DMD's RAM, which significantly reduces the time for data transfer. Different sets of random patterns are employed across various experiments to minimize the likelihood of accidentally favorable patterns and to demonstrate the robustness and consistency of 2P-FOCUS. 2P-FOCUS effectively improves signal intensity by leveraging the nonlinearity of multiple-beam interference and two-photon excitation. Unlike phase modulation, intensity modulation selects multiple in-phase beams that constructively interfere after passing through scattering media. The nonlinearity of multiple-beam interference greatly enhances the intensity in the central lobe compared to the linear superposition of multiple incoherent beams. Two-photon excitation further amplifies this nonlinearity compared to one-photon excitation. Overall, the fluorescence intensity is proportional to the $4^{th}$ power of the multiple-beam field in the Fourier domain, making the fluorescence intensity extremely sensitive to the intensity modulation. Although modulating the intensity of light at the Fourier plane can induce a phase change at the image plane (as in Lee holography), 2P-FOCUS calculates the intensity mask directly from the fluorescence intensity of multiple-beam interference rather than relying on phase correction using Zernike modes. This approach classifies our method as an intensity modulation technique rather than a phase modulation technique. 2P-FOCUS also enables subregion correction and highlights the importance of subregion correction for addressing scattering beyond the memory effect range. We demonstrated 2P-FOCUS by correcting scattering in mouse brain tissue ex vivo over a 230$\times$230$\times$510 $\mu m^3$ volume, successfully improving the fluorescent intensity by 36.1-fold after correction. 

To apply 2P-FOCUS for in vivo imaging in the future, the current 2P-FOCUS needs to be improved in two key aspects. First, to extend the imaging depth further, 2P-FOCUS needs to acquire high signal-to-noise ratio measurements even deep inside the sample. This proves challenging for the current version of 2P-FOCUS, as measurements are taken under completely random modulation in the Fourier domain. Sufficient signal-to-noise ratio of measurements is necessary for calculating an accurate correction mask, as shown in Fig.\ref{fig2}J. This limitation could potentially be overcome by incorporating the correction mask from the plane above to generate the correction mask for the current plane. Second, the current process of generating a correction mask in 2P-FOCUS is not fast enough to correct millisecond-scale dynamic scattering due to blood flow \cite{Qureshi2017-on}. The current correction speed may be slow for imaging through blood flow, but it should be sufficient for imaging areas without arteries, as previous works \cite{Papadopoulos2016-sy} based on wavefront shaping show that the correction mask can last for tens of minutes in vivo. The maximum patterning rate of the DMD is 12.5 kHz, which means the correction process could take less than 1 s if the DMD could be operated at full speed with upgraded electronics. Therefore, the correction speed of 2P-FOCUS could be further improved with a higher bandwidth data acquisition card, a higher repetition rate laser, and software designed for auto-correction with closed-loop control.

In summary, 2P-FOCUS is the first two-photon microscopy system to use intensity modulation for rapid and effective scattering correction. It also uses random patterns as the orthogonal basis and employs a single-shot algorithm to complete the correction within a few seconds. This method can significantly boost fluorescence intensity by several tens of times at the same illumination power on the sample. This capability makes it a powerful tool for deep tissue imaging, facilitating research in neuroscience, immunology, and cancer.

\begin{backmatter}
\bmsection{Funding}
Research reported in this publication was supported by the National Institute Of General Medical Sciences of the National Institutes of Health under Award Number R35GM155193. The content is solely the responsibility of the authors and does not necessarily represent the official views of the National Institutes of Health. This work was also supported by Dr. Yi Xue’s startup funds from the Department of Biomedical Engineering at the University of California, Davis. 

\bmsection{Acknowledgments}
We thank Hillel Adesnik and Janine Beyer at the University of California, Berkeley, for preparing the transgenic mouse brain in Fig. 4 and for providing the PMT used in the microscopy setup. We thank Jiandi Wan and Brianna Urbina at UC Davis for preparing the mouse brain in Fig. 5. We thank Weijian Yang and Zixiao Zhang at UC Davis for helping us measure the pulse width of the laser light before and after dispersion compensation. We thank Peter T. C. So at the Massachusetts Institute of Technology and Laura Marcu at the University of California, Davis, for proofreading the manuscript.

\bmsection{Disclosures}
The authors declare no conflicts of interest.

\bmsection{Data Availability Statement}
The data that support the plots within this paper and other finds of this study are available from the corresponding author upon reasonable request. 

\bmsection{Supplemental Visualization}
Supplemental Visualization 1 shows the 3D image stack of neurons over 450 $\mu$m depth. Supplemental Visualization 2 shows the 3D iamge stack of blood vessels over 510 $\mu$m depth. 

\end{backmatter}

\begin{appendices}
\section*{Appendices}
\setcounter{figure}{0} 
\renewcommand{\thefigure}{S\arabic{figure}}
\hypertarget{appendixdoc}{}
\section{2P-FOCUS Setup}\label{setup}
The laser source for 2P-FOCUS is a femtosecond pulsed laser at 1035 nm wavelength and 1 MHz repetition rate (Monaco 1035-40-40 LX, Coherent). The maximum power used for 2P-FOCUS is 2.4 W (The maximum total power of the laser is 40 W but 37.6 W is used to pump an optical parametric amplifier, which is not used for 2P-FOCUS). A polarizing beam splitter cube (PBS123, Thorlabs) and a half-wave-plate (WPHSM05-1310) mounted on a rotation mount (PRM05, Thorlabs) are used to manually adjust the input power to the following optics in the system. A 2-axis galvo-mirror system (GVS002, Thorlabs) are placed on the Fourier plane to scan the laser beam in 2D for imaging in Fig. 4 and Fig. 5. The laser beam is then expended with a 4-$f$ system (L1, LA1401-B, $f_1$ = 60 mm, Thorlabs; L2, AC508-100-C-ML, $f_2$ = 100 mm, Thorlabs). Next, the laser beam is pre-dispersed by a ruled grating (GR13-0310, 300/mm, 1000 nm blaze, Thorlabs) on the Fourier plane to compensate for the dispersion induced by the DMD. After the grating, the pre-dispersed beam is relayed to the DMD with a 4-$f$ system (L3, LB1199-C, $f_3$ = 200 mm, Thorlabs; L4, AC508-200-C-ML, $f_4$ = 200 mm, Thorlabs). The DMD (DLP650LNIR, 1280$\times$800 pixels, maximum pattern rate 12.5 kHz, VIALUX) is placed on the Fourier plane to project binary intensity masks. The beam from the DMD is relayed to the back aperture of the objective lens by two 4-$f$ systems, which is simplified as one 4-f system in the optical schematic diagram in Fig. 1D. The first 4-$f$ system is to expend the beam (AC508-150-C-ML, $f$ = 150 mm, and LA1256-C, $f$ = 300 mm). The second 4-$f$ system is to 1:1 relay the beam (two AC508-200-C-ML, $f$ = 200 mm). A dichroic mirror (FF880-SDi01-t3-35x52, Semrock) is used to reflect the beam to the back-aperture of the objective lens (XLUMPlanFL N, 20$\times$, 1.00 NA, water immersion, Olympus). Samples are placed on a manual 3-axis translation stage (MDT616, Thorlabs). In the emission path, a shortpass filter (ET750sp-2p8, CHROMA) is used to block the reflected excitation light. A bandpass filter (AT635/60m, CHROMA) is used in the experiments for Fig. 2 and Fig. 4 to pass through red fluorescence, while it is removed in the experiments for Fig. 5 because the emission wavelength of FITC-dextran is 520 nm. The fluorescence is detected by a PMT (H15460, Hamamatsu). A sCMOS camera (Kinetix22, Teledyne Photometrics) is used to visualize the focus in Fig. 2F-H. A beam turning cube (DFM1-E02, Thorlabs) is used to switch between the camera and the PMT. In addition, a one-photon widefield microscope is implemented, overlapping with 2P-FOCUS, to locate the sample and find the focal plane before two-photon imaging. The one-photon system consists of a LED (M565L3, Thorlabs), an aspherical condenser lens (ACL25416U-A) to collimate the LED light, and a dichroic mirror (AT600dc, CHROMA) to combine the one-photon path to the two-photon path. A power meter (PowerMax USB - PM10-19C Power, Coherent) is used to measure laser power. During imaging sections, the illumination power on the sample is measured at a relayed image plane located before the objective lens but after the DMD. This measurement is then multiplied by the power loss rate due to the optical parts situated between the relayed image plane and the sample. 

The 2P-FOCUS system is controlled by a computer (OptiPlex 5000 Tower, Dell) using MATLAB and a data acquisition card (PCIe-6363, X series DAQ, National Instruments) for signal input/output. Voltage signals for externally triggering the laser and the DMD, as well as for controlling the scanning location, are generated using MATLAB and delivered to the devices by the DAQ card. Simultaneously, the fluorescence intensity detected by the PMT is read by the DAQ card through an analog input port. 

\section{Dispersion Control}\label{Dispersion}
Dispersion control has been implemented when using a DMD to modulate femtosecond laser light\cite{Cheng2015-jy}. The pixels on the DMD tilt to $+12^\circ$ or $-12^\circ$ along the diagonal direction of the pixels when projecting binary patterns, thereby inducing dispersion to the input excitation beam like a blazed reflective grating. This process can be described by the grating equation:
\begin{equation}
\Delta = d(sin\theta_i+sin\theta_m) = m\lambda,\label{eq10}
\end{equation}
where $\Delta$ denotes optical path difference, $d$ denotes the spacing between grooves, $\theta_i$ denotes the incident angle, $\theta_m$ denotes the diffraction angle, $m$ is the order of principal maxima, and $\lambda$ denotes the wavelength of light ($\lambda = 1035\pm5$ nm). The effective spacing between grooves of the DMD is calculated when the DMD is operated in the Littow configuration, that is, $\theta_i = \theta_m = 12^\circ$ and $m = 1$. Therefore, the effective spacing between grooves of the DMD, $d_{DMD}$, is 2.49 $\mu$m, which is different from the pitch size (10.8 $\mu$m) of the DMD. 

To compensate the angular dispersion induced by the DMD ($\mathfrak{D}_{DMD}$), we implemented a grating on the conjugate Fourier plane with a 4-f relay system (L3, L4 in Fig.~\ref{fig1}D) to the DMD. The angular dispersion induced by the grating, $\mathfrak{D}_G$, satisfies:
\begin{equation}
\frac{\mathfrak{D}_G}{\mathfrak{D}_{DMD}} = \frac{f_4}{f_3},\label{eq11}
\end{equation}
where the anglular dispersion is calculated as
\begin{equation}
\mathfrak{D} = \frac{m}{d \cos\theta_m}.\label{eq12}
\end{equation}
$f_3$ and $f_3$ are the focal length of lens L3 and L4, respectively. Considering the effective grooves ($1000/2.49 = 402$ grooves/mm) and the size of the DMD (13.8 $\times$ 8.6 $mm^2$), a grating with 300 grooves/mm and a 1:1 relay system are selected. The output beam from the DMD is designed to be perpendicular to the DMD, that is, $\theta_m = 0$. According to Eq. \ref{eq10}, \ref{eq11}, \ref{eq12}, when $m = 1$, the dispersion angle of the incident beam ($\lambda = 1035\pm5$ nm) is $0.25^\circ$. Correspondingly, when $f_3 = f_4 = 200$ mm and $m = 1$, the incident angle to the grating is $25.1^\circ$. After the dispersion control, the focus becomes a symmetric circular spot (Fig.~\ref{fig1}E).

\section{Data Processing}
\subsection{Process data acquired by the PMT.} 
Data are acquired for 1 ms per random pattern using the PMT in the first step of scattering correction. Within the 1 ms time frame, the laser is turned on for 0.9 ms and off for 0.1 ms. The measurements acquired while the laser is on are subtracted from the measurements acquired while the laser is off. After subtraction, all negative voltages are set to zero. Next, a box filter with a width of 3 pixels is applied to the time-lapse signal to remove fluctuations. The final value of the PMT signal for one random pattern modulation is the sum of the processed time-lapse voltage signals acquired over 0.9 ms. Repeating this process for PMT data acquired under all random patterns generates Fig. 2B.

\subsection{Generate a correction mask from the processed PMT data.} The processed PMT data is sorted, and the random patterns corresponding to the top 10\% brightest data are selected. The sum of the selected random patterns generates Fig. 2C.

\subsection{Binarize the grayscale correction mask.} The intensity distribution on the DMD is measured with a fluorescence slide without scattering media before every experiment, and is referred to as an intensity calibration map $I_{map}(f_x, f_y)$. The intensity calibration map is measured by turning on one super-pixel at a time and recording the corresponding fluorescence intensity. The laser power at the image plane, $P_0$, is also measured when all pixels are turned on. Therefore, the laser power $P$ of any intensity mask $M(f_x, f_y)$ (a random mask or a correction mask) can be calculated by $P = P_0\sum_{f_x = 1}^{800} \sum_{f_y = 1}^{800} M(f_x, f_y)\cdot I_{map}(f_x, f_y)$. By adjusting $P_0$ and/or $M(f_x, f_y)$, the laser power on the sample $P$ is kept the same before and after correction. The largest improvement is achieved when $P_0$ is maximized and $M(f_x, f_y)$ is highly selective. In this case, $P_0$ is measured when the input power to the DMD is maximized by turning the half-wave plate. With the known values of $P_0$, $I_{map}$, and $P$, the threshold for binarizing the grayscale correction mask is calculated accordingly. 

\subsection{Generate and process 2D images} An image acquired by the PMT (Fig. 3---5) is generated by reassign the processed PMT data to their corresponding scanning locations. The first step is to remove the background from the image by subtracting the mean of 300 pixels with the lowest intensity, and setting all negative intensities to zero. Next, a 2$\times$2 median filter is applied to the image to remove salt-and-pepper noise. 

For images acquired by the camera (Fig. 2F-G), the background is removed by subtracting the images with a background image, and then a 3$\times$3 median filter is applied to remove noise.  

\subsection{Generate the volumetric view of 3D image stack.} Fig. 4A and Fig. 5A are generated using ImarisViewer 10.1.0. The 3D image stacks consist of processed 2D images. The image stack is interpolated by 10 times along the $z$-axis for a better 3D display.

\section{Sample Preparation}\label{sample}

\subsection{Homogeneous fluorescence slide with bone for Fig. 2} A microscope slide (CAT. NO. 3049, Gold Seal) is coated with a thin layer of fluorescent paint (Tamiya color, fluorescent red). After the paint is dry, a piece of chicken bone is glued on it as the scattering medium using clear gorilla glue.

\subsection{Fluorescence beads with bone for Fig. 3} Red fluorescence beads suspension (R700, Thermo Fisher Scientific, MA) is mixed with PDMS (Sylgard 184, Dow Inc, MI) in the ratio of 1:550. Base elastomer and curing agent of PDMS are mixed in the ratio of 10:1. After mixing, we use a vacuum desiccator to remove the air bubbles in the mixture for 30min. Then the mixture is poured onto a clean microscopy slide, covered by a coverslip, and heated at 100 Celsius for 35min to cure the PDMS. After the fluorescent beads are fixed, a piece of chicken bone is glued on top of it as the scattering medium using clear gorilla glue.

\subsection{Whole brain preparation for Fig. 4}The PV tdTomato mouse (PV-IRES-Cre;LSL-tdTomato (Ai9)) was weighed and put into 5\% isoflurane for initial induction. An IP injection of Ketamine hydrochloride 40-80 mg/kg + Xylazine 5-10 mg/kg was given and then the animal was put back into the isoflurane until the animal's breathing ceased. The animal was brought to the chemical fume hood and placed in the supine position. An additional amount of isoflurane was placed in a 10 cc syringe with a gauze and placed over the mouse’s nose for additional anesthesia. The hair on the ventral thorax was soaked with 70\% alcohol. The anesthetic depth was checked via lack of toe pinch response. A midline incision was made through the skin over the proximal abdomen and thorax. The skin was dissected to expose all underlying muscle. A cut was made into the abdomen, the diaphragm was punctured and  a thoracotomy was made by bilateral para- midline incisions through the ribs toward the thoracic inlet, exposing the thoracic viscera. The catheter was placed in the left ventricle, the right atrium was cut to exsanguinate the mouse and allow for drainage of the perfusate. The animal was perfused with cold 4\% paraformaldehyde, approximately 12 mls. The brain was dissected from the skull and then placed in cold 4\% paraformaldehyde.

\subsection{Fluorescein isothiocyanate-dextran (FITC-dextran) injection and whole tissue preparation for Fig. 5} Anesthesia was induced in mice with 2.0-3.0\% isoflurane and maintained at 1.5-2.0\%. Depth of anesthesia was monitored by toe-pinch and body temperature was maintained by a water perfused thermal pad (Gaymar T/Pump) set at 37$^\circ$C. Fluorescein isothiocyanate-dextran (FITC-dextran) (MW = 2 MDa; Sigma-Aldrich) was injected retro-orbitally at a total volume of 150 $\mu$L. After retro-orbital injection, FITC-dextran was allowed to circulate for a total of 10-15 minutes and the animal was sacrificed. After the animal was sacrificed, the whole brain was collected and stored overnight in phosphate-buffered saline at 4$^\circ$C and was imaged the next day.

\section{Supplementary Tables and Figures}
\hypertarget{appendixtable}{}
\setcounter{figure}{0} 
\begin{figure}[ht!]
\centering
\includegraphics[width=1\textwidth]{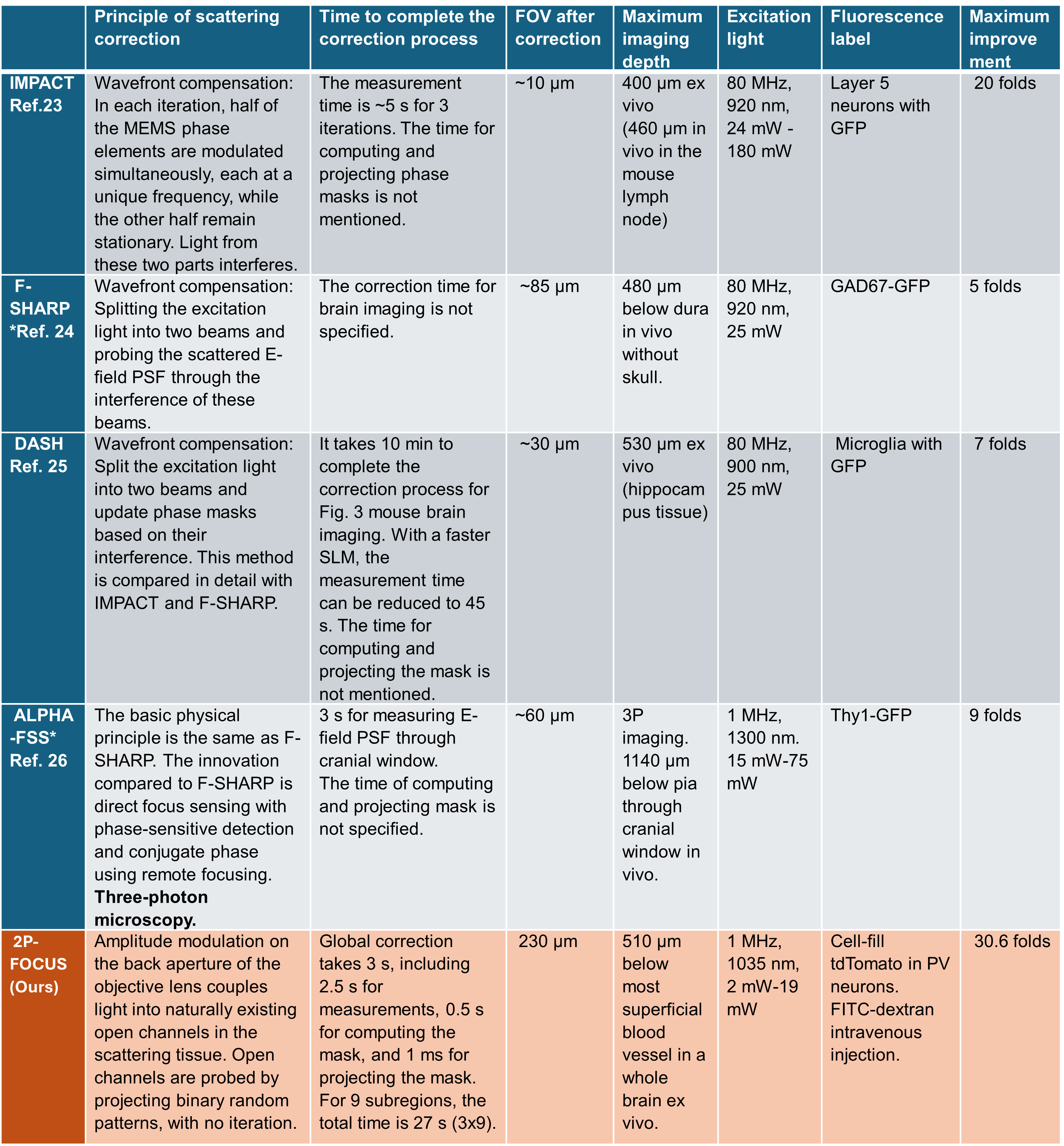}
\caption{Table S1. Comparison between state-of-the-art multiphoton microscopy systems using active scattering correction and the 2P-FOCUS system. The reference numbers in the table refer to the bibliography in the main text.}\label{table1}
\end{figure}

All methods are based on two-photon microscopy except ALPHA-FSS \cite{Qin2022-ns}, which is a three-photon microscopy system. The parameters listed in the table are collected from mouse brain images. $*$ These two works also demonstrate imaging through the skull in vivo, in addition to imaging through a cranial window. When imaging through the skull, F-SHARP \cite{Papadopoulos2016-sy} reaches a depth of 325 $\mu$m through a 50 $\mu$m-thick skull, and ALPHA-FSS \cite{Qin2022-ns} reaches a depth of 780 $\mu$m through a 100 $\mu$m-thick skull.

\setcounter{figure}{0} 
\renewcommand{\thefigure}{S\arabic{figure}}

\begin{figure}[h!]%
\centering
\includegraphics[width=1\textwidth]{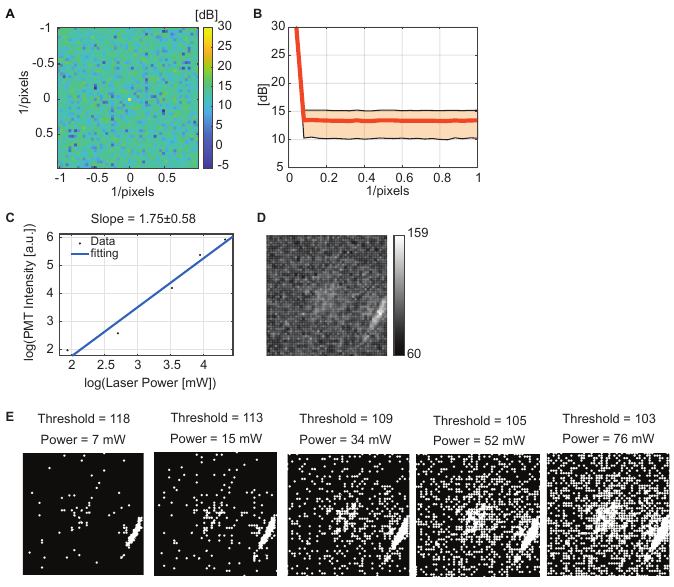}
\caption{The influence of the binarization threshold on correction masks. (A-B) Frequency analysis of 2,500 random patterns with a super-pixel size of 8 pixels and a sparsity of 0.4. (A) The 2D frequency spectrum of a single random pattern from the set. (B) The average frequency spectrum along the x-axis across the 2,500 random patterns. The red line represents the mean value, and the orange shading indicates the standard deviation. (C) Log-log plot of the fluorescence intensity before correction as a function of the laser power on the sample. The slope of the log-log plot is 1.75$\pm$0.58, indicating that the fluorescence intensity before correction increases quadratically with the illumination power. (D) The grayscale correction mask, the same as in Figure 2C. (E) The binary correction masks generated by applying different thresholds to (B). These masks are used to produce the data for ``after correction" in Figure 2I.}\label{figS1}
\end{figure}

\begin{figure}[ht!]%
\centering
\includegraphics[width=1\textwidth]{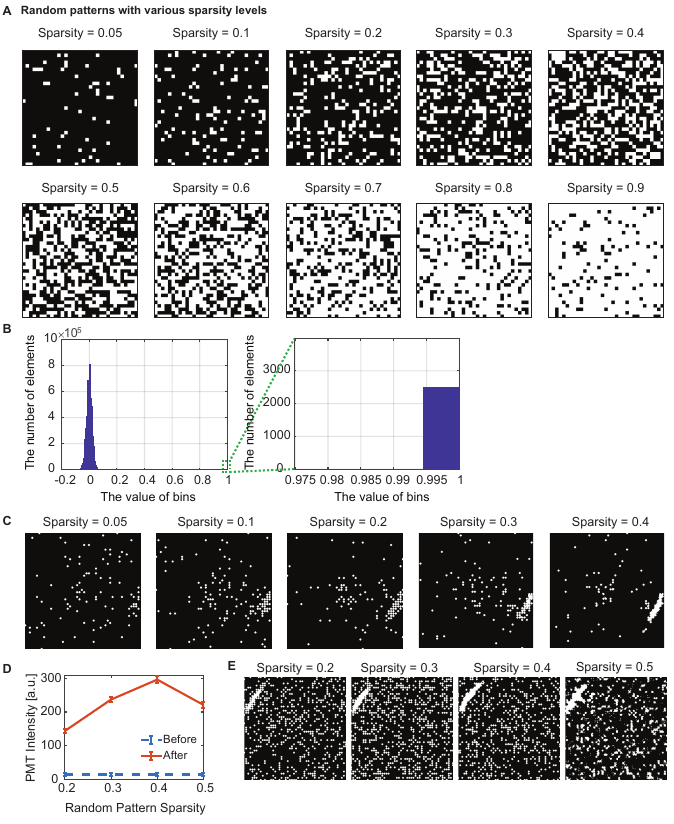}
\caption{The influence of the sparsity of random patterns on correction masks. (A) Random patterns with varying sparsity levels. Sparsity refers to the percentage of pixels that are turned on out of the total number. When the sparsity is between 0.5 and 0.9, the random patterns act more like notch filters, determining which pixels should be turned off rather than on. (B) Histogram of all elements in the matrix $I = P^TP$ (left) and a zoomed-in view (right). The zoomed-in view highlights the values of the diagonal elements in matrix $I$, while the other bins represent the non-diagonal elements. The results indicate that the dot product of any two distinct vectors in matrix $P$ is $0\pm0.02$, and the dot product of a vector with itself is 1.} (C) Five correction masks used to generate the ``after correction" data in Figure 2J. (D) Fluorescence intensity before (blue dashed line) and after (red line) correction as a function of random pattern sparsity. Data were collected by focusing a beam through another region of the bone. The results show that the fluorescence intensity decreases when the correction mask is generated with random patterns having a sparsity of 0.5 compared to those with a sparsity of 0.4. (E) Four correction masks used to generate the ``after correction" data in (D) from random patterns with various sparsity levels.\label{figS2}
\end{figure}

\setcounter{table}{1} 
\renewcommand{\thetable}{S\arabic{table}}

\begin{table}[ht!]
\centering
\begin{tabular}{||c|c|c|c|c|c|c||} 
 \hline
  Fig. 2J & Baseline & Sparsity 0.05 & Sparsity 0.1 & Sparsity 0.2 & Sparsity 0.3 & Sparsity 0.4\\ [0.5ex] 
 \hline\hline
 Mean & 7.21 & 12.02 & 66.24 & 101.15 & 191.61 & 331.23 \\ 
 \hline
 STD & 0.75 & 1.22 & 3.28 & 3.72 & 5.29 & 7.29 \\
 \hline
Ratio & 1 & 1.67 & 9.19 & 14.03 & 26.58	& 45.95 \\ [1ex] 
 \hline
 \end{tabular}
\caption{The dataset from Fig. 2J and the corresponding improvement ratios.}
\label{TableS1}
\end{table}

\begin{table}[ht!]
\centering
\begin{tabular}{||c|c|c|c|c|c||} 
 \hline
  Fig. S2D & Baseline & Sparsity 0.2 & Sparsity 0.3 & Sparsity 0.4 & Sparsity 0.5\\ [0.5ex] 
 \hline\hline
 Mean & 14.42 & 143.68 & 236.89 & 296.05 & 221.11 \\ 
 \hline
 STD & 3.26 & 6.28 & 7.56 & 10.99 & 8.80 \\
 \hline
Ratio & 1 & 9.97 & 16.43 & 20.54 & 15.34 \\ [1ex] 
 \hline
 \end{tabular}
\caption{The dataset from Fig. S2D and the corresponding improvement ratios.}
\label{TableS2}
\end{table}

\begin{figure}[ht!]%
\centering
\includegraphics[width=1\textwidth]{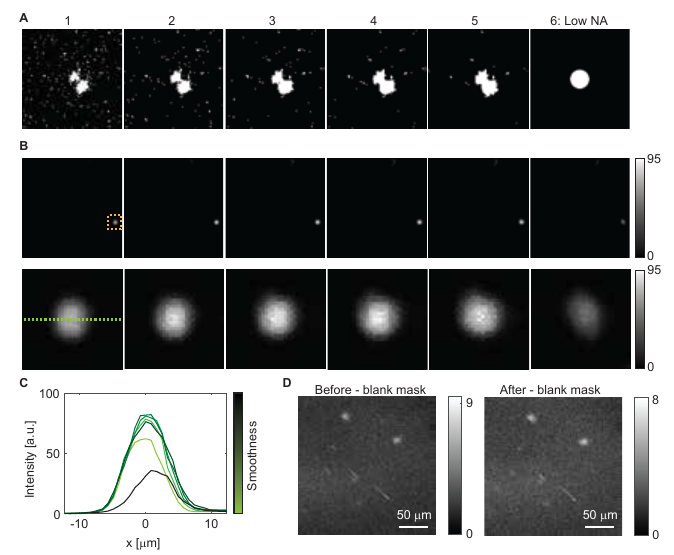}
\caption{(A) Correction masks with varying degrees of smoothing (plots 1–5), as well as a low-NA mask (plot 6). These masks provide the same output power given the same input power to the DMD. (B) Imaging of a red fluorescent bead through a piece of chicken bone under the intensity modulation of the corresponding masks. The second row shows a zoomed-in view of the region highlighted by the yellow box in the first row. All images are displayed using the same color scale. The maximum intensity is 95, achieved with the 4th correction mask, while the peak intensity under the low-NA mask is 42. (C) Intensity profile of the cross-section along the green dashed line in (B). (D)} Examining photobleaching due to correction processes. The peak intensity decreased from 9 to 8 when comparing the image taken before the process without correction and the image taken after the process without correction.\label{figS3}
\end{figure}

\begin{figure}[ht]%
\centering
\includegraphics[width=1\textwidth]{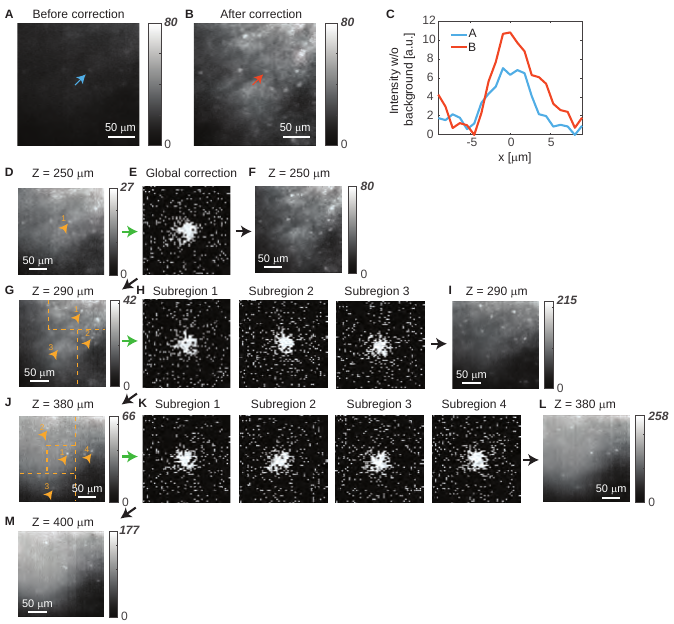}
\caption{Scattering correction for imaging PV neurons deep in the mouse brain. (A-C) Comparison of the images taken (A) before and (B) after global correction under the same color bar. The images are the same as in Figure 4E-F. The neuron used as a reference (pointed out by the arrows) becomes brighter rather than dimmer after correction. (C) The intensity profile of the reference neuron before and after correction. Background intensity is subtracted. The peak intensity of this neuron is improved by about 1.8-fold, which is not as much as the brightest neurons. (D-M) All correction performed for taking the $0-450 \mu m$ image stack in Figure 4. The green arrows indicate the image is used to produce the corresponding mask. The black arrows point at images after correction. The yellow arrows in the images point to fluorescence objects used as references. Notice that (G), (J), and (M) are taken with last correction mask rather than blank screen on the DMD. (D-F) The first scattering correction is performed at 250 $\mu$m depth. (D) The image before correction, which is the same as Figure 4E. (E) A single correction mask is produced, which is the same as the bottom plot in Figure 4D. (F) The image after correction, which is the same as Figure 4F. (G-I) The second scattering correction is performed at 290 $\mu$m depth. (G) Image taken with the global correction mask in (E). The field-of-view is divided into three subregions. Notice that we used the background fluorescence as the reference in the third subregion to generate the correction mask. (H) Three correction masks corresponding to the three subregions. (I) Image after subregion correction. (J-L) The third scattering correction is performed at 380 $\mu$m depth. (J) Image taken with the three correction masks in (H). Four neurons are identified as the references for four subregions. (K) Four correction masks (identical to Figure 4J) corresponding to the four subregions. (L) Image after correction. (M) Image taken with the four correction masks (identical to Figure 4I) in (K) at 400 $\mu$m depth. }\label{figS4}
\end{figure}

\begin{figure}[ht]%
\centering
\includegraphics[width=1\textwidth]{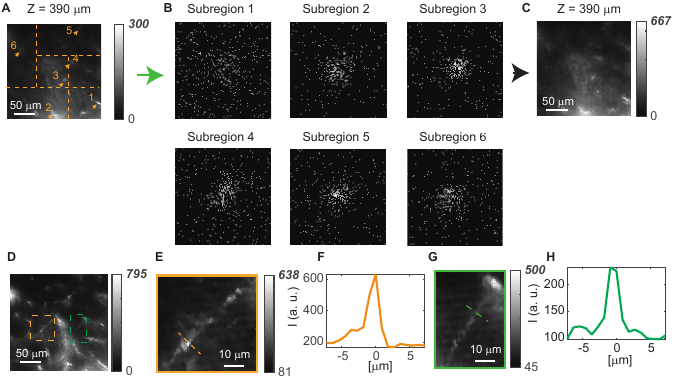}
\caption{Subregion correction is performed at 390 $\mu$m depth. (A) Image taken with the four correction masks in Fig.  5, identifying six fractions of blood vessels as references for six subregions. (B) Six correction masks corresponding to the six subregions. (C) Image after correction. (D) The same as Fig.  5E. (E, G) Zoomed-in views of capillaries in the boxed regions in (D). (F, H) Intensity profiles of the cross-sections of the capillaries in (E) and (G), marked by the dashed lines. Fine capillaries are resolvable after applying the correction masks. }\label{figS5}
\end{figure}

\end{appendices}
\clearpage
\bibliography{FocusBibliography}

\end{document}